\def\beg{\begin{equation}}
\def\eeq{\end{equation}}
\documentstyle[12pt]{article}
\begin{document}
\begin{center}
{\Large{\bf Comments on ``Composite Fermion (CF) model of 
quantum Hall effect - Two-dimensional electron system in 
high magnetic fields, S. S. Mandal, M. R. Peterson and 
J. K. Jain, Phys. Rev. Lett. {\bf 90}, 106403 (2003).}}
\vskip0.45cm
{\bf Keshav N. Shrivastava}
\vskip0.25cm
{\it School of Physics, University of Hyderabad,\\
Hyderabad  500046, India}
\end{center}

 The flux quanta attachment to the electrons creates composite 
fermions (CFs). The mass, the size and the density of the CF are 
inconsistent with real material. The sequence of fractional 
charges which suggest formation of CF agrees with the data but 
there are no monopoles in GaAs. Hence the CF model is internally 
inconsistent. 
There are two options at this stage. (1) The flux quanta are 
attached 
to the electrons. This is a theoretical  possibility with nothing 
to 
do with any experiment ever performed within the last seventy five 
years. It will violate Maxwell equations and create unreal objects.
(2) The CF give a sequence which is deduced from the experimental 
data and hence agrees with the data. In this case, the masses, the 
sizes and
 the densities are internally inconsistent. 

\vskip0.25cm
Corresponding author: keshav@mailaps.org\\
Fax: +91-402-301 0145.Phone: 301 0811.
\vskip0.25cm
\noindent {\bf 1.~ Introduction}

     The paper by Mandal et al shows that even number of magnetic flux
 quanta are attached to the electron. We find that this flux attachment 
is internally inconsistent. If there  is a contact between the sequence 
and the data then the model does not have Lorentz invariance and mass 
and size are inconsistent. If the flux quanta are attached to the 
electron then there is no contact of this flux-attached model with the
 experimental data on quantum Hall effect because there are no monopoles 
in GaAs. By using the experimental numbers, Jain[1-3] constructed a
 sequence which gives the correct plateaus in the quantum Hall effect.
 From these sequences, effort is made to construct a formula for the 
magnetic field. This field formula is interpreted in terms of flux
 quanta attachment to the electron but it is internally inconsistent. 

      We find that whenever effort is made to compare this composite 
fermion (CF) model with the data, some thing or the other is always 
wrong. There are many different number of flux quanta required to
 reach agreement with the experimental data. These are denoted by
 $^6$CF, $^8$CF and $^{10}$CF.

     Several questions arise. This CF model is correct and should be 
accepted as new physics. It will be immediately obvious that CF model 
is not internally consistent and hence should be discarded. Let us
 examine a few points.

\noindent{\bf 2.~~Comments.}

{\it (i) Abundance of flux quanta and Monopoles}

     Usually the flux is generated by the flow of current in a coil.
 In the CF model, the current which generates the flux quanta is missing. On page 106403-2, left column, para 2, it is mentioned that

 ``N electrons are confined to the two dimensional surface of a sphere,
 moving under the influence of a radial magnetic field produced by a
 magnetic monopole of strength, Q, at the center. According to Dirac's 
quantization condition, Q can be either an integer or a half integer, 
and produces a total flux of 2Q$\phi_o$ ..." [Surely, Dirac's monopoles
 do not exist in GaAs/AlGaAs and hence the CF model does not apply]. 
Dirac symmetrized the Maxwell equations and predicted the monopoles 
in 1931 but these monopoles have not been seen in any of the experiments. Although 'tHooft[2] removed the string of the Dirac's monopole, it is still not found in nature.  In the 'tHooft's theory, the Maxwell equations remain unchanged except at one point. Therefore, whereas 
there are no monopoles
in the Maxwell theory, there is a monopole in 'tHooft's. There is a 
report of finding a monopole, the mass of which may be equal to that 
of an atom of atomic number ~200. In that case, the difference between
 a heavy atom and a monopole was not resolved. In that case the mass
 of the monopole will be about 10$^5$ times larger than that of an 
electron. Naturally impossible in GaAs . The number of such monopoles 
will be very small but $^8$CF or $^{10}$CF require far too many
 monopoles to be present. As far as the number of monopoles is 
concerned for $^{10}$CF, 10 times more monopoles are required than
 the  number of electrons and what will be the mass of such a large 
number of 
monopoles? It is obvious that the CF model is demanding far too many 
flux quanta than can be available. Therefore, on the grounds of 
abundance of free flux quanta, the CF model can not be accepted as a 
new theory.

{\it(ii) Mass.}

     What is the mass of a vortex? The vortices are found in 
superconductors with unit flux $\phi_o$=hc/2e. In the normal state we 
need not take the charge as $2e$ but $e$. Hence the normal unit flux
is $\phi_o$=hc/e. The mass was calculated by Suhl and found to be
 10$^4$ per unit length in the units of the mass of the electron. If 
we take an order of magnitude of a flux mass as 10$^4$, then no such 
free mass is available in GaAs. If the CF model is correct, then the
 mass of GaAs after CF formation will become 10$^4$ times the mass
 before CF formation. Therefore, CF model is surely inapplicable. Let 
us examine this mass from another view point. The magnetic length is 
$l_o$=$(\phi_o/B)^{1/2}$. At 8 Tesla, the magnetic length is about 
10$^{-6}$ cm so that the wave vector is 10$^{6}$ cm$^{-1}$. The mass
 due to kinetic energy is the mass per unit length ~mk$_F \sim m\times 
10^{6}$.
The wave vector of the electron is about 10$^{8}$ cm$^{-1}$ so that for
 the same energy, which is of the order of cyclotron energy, the mass
 of the vortex should be 100 times larger than the electron mass. 
Therefore, when $^{10}$CF are formed the mass of the CF will be 1000 
times the mass of an electron but there is no evidence of such a large 
massive quasiparticle.
    
{\it (iii) Free flux.}

     Let us take a positive approach to the problem. The fluxes that 
get attached to the electrons in any of the various composite fermion
 models are not physical magnetic fields and therefore need no
 persistent currents to create them. They are either short hand for 
rearrangement
of purely electronic states or a mathematical transformation that does
 nothing but change the description of the system but no such 
transformation is known. The general motivation for the 
``attached flux"
language is that one imagines putting a real magnetic solenoid through 
a point (or electron) in the 2DEG and gradually increases the field in 
the solenoid untill it contains one unit of flux, (it may never happen
 so).
The electric field due to the changing magnetic field acts on the other
 electrons outside the solenoid and changes their state. The final unit
 flux, however, is invisible to all electrons since only fractional
 fluxes produce a Bohm-Aharonov effect. It is therefore discarded and 
the word ``attached flux" is retained as the appropriate language to
 describe the rearranged many-electron state created by this mechanism.
[Jain does not do this any way and this mechanism will not produce the
experimental sequence of plateaus]. In a Chern-Simon field, a unit 
flux does nothing physical at all. It allows us to describe the 
Laughlin state as a kind of Bose condensate of the purely mathematical
 composite boson.

     A good analogy for these purely mathematical transformations is
 provided by the two dimensional Ising model. A Jordan-Wigner
 transformation which is similar in character to a ``flux attachment",
 turns the spins into Majorana fermions.(Majorana means that which
 exchanges the space coordinates of two interacting particles). These 
are free fermions so the model is easily solved in this language. All
 these ``attached fluxex" are loosely related to the operation but
 all have  in common that no real magnetic flux is being attached to
 anything. Similarly, Holstein-Primakopf transformation transforms
 the spin operators into bosons
so that the boson Hamiltonian can be solved by usual methods.
The fermion transformation does not introduce internal inconsistencies
nor does the boson transformation. In Jain's CF there are too many
 inconsistencies and the whole thing does not work like a transformation.
Therefore, even if the ``flux attachment" model is correct, the series
 which agree with the quantum Hall effect data are unphysical. Hence
 when ``flux attachment" models are solved, they will have nothing to
 do with the experimental data. In the case of Chern-Simons fields,
 the scalar potential term has been set to zero and then only the
 vector potential is changed so there is no Lorentz condition. 
Therefore, Jain's CF model has nothing to do with the experimental 
work on GaAs. The electron scattering or attachment to magnetic 
monopoles may be interesting by itself as a good problem but it will
 have nothing to do with quantum Hall effect. Dyakonov[5] and 
Farid[6] have also shown that there is a lack of theoretical basis in
 CF model.

{\it (iv) Exactness.}

     Some of the parts of the calculation are exact but Laughlin has 
used $e$ in place of $e/a_o^2$ where $Ba_o^2$=$\phi_o$. Therefore, $e$
becomes fractional or $B$, cannot be resolved. Laughlin performed the 
calculation of energy for $m$=1/3 and 1/5 but suggested that the ground
 state may cross the charge-density wave at 1/7. Now, Mandal et al[3]
 have extended the calculation to 1/7, 1/9 and 1/11 and exactness is
 preserved. This means that $^6$CF, $^8$CF and $^{10}$CFs are correct
 but attaching 10 flux quanta to one electron has no chances of
 agreeing with size or mass requirements and the density of $^{10}$CFs, 
on physical grounds can not be equal to that of electrons and hence 
CF model becomes internally inconsistent. Is there enough room in GaAs
 to keep all these flux quanta? The answer is, no. There is no room 
to keep so many flux quanta. So what is an exact answer is also a wrong 
answer. So if there is a model of flux attachment to electrons, then it 
will have nothing to do with the experimental data. If the series 
agrees with the data, then it has nothing to do with flux attachment.

\noindent{\bf3.~~ Conclusions}.

   The attachment of flux quanta can work like a Holstein-Primakopf
 transformation but Jain has not found the transformation. As and when
 such a transformation is found, it will not be a theory of GaAs like
 real material. The sequences which are made to agree with the 
experimental data create internal inconsistencies. The exactness is 
limited to the product of charge and the field and hence can not 
resolve whether charge or field has been fractionalized. The 
experimental data is not consistent with 1/odd fractions. It is
 important to learn the numerator also. The mass of the vortices is
 not consistent with GaAs.
The abundance of the vortices is also not consistent with what is 
possible in GaAs. Pan et al[7] have also noted that the experimental 
data are not consistent with the sequences suggested by Jain[1-3].
There are too many inconsistencies in the CF model and hence they can
 not be resolved. If the flux attachment transformation is worked out
it will not be relevent to experimental work. Hence, CF model should 
be discarded in as much as its comparison with the experimental data
 is concerned [8].

\noindent{\bf About the author}: {\it Keshav Shrivastava has obtained 
Ph.D. degree from the
Indian Institute of Technology and D. Sc. from Calcutta University. 
He is a member of the American Physical Society, Fellow of the 
Institute of Physics (U.K.)and Fellow of the National Academy of 
Sciences, India. He has worked in the Harvard University, University
 of California at Santa Barbara, the University of Houston and the 
Royal Institute of Technology Stockholm. He has published 170 papers 
in the last 40 years. He is the author of two books. Shrivastava's 
paper with Roy Anderson,
published in J. Chem. Phys. 48, 4599(1967) was found to be useful 
by M. C. R. Symons,F. R. S. The letter with K. W. H. Stevens,
 J. Phys. C 3, L 64 (1970) was useful to D. I. Bolef, the paper
 with Vincent Jaccarino,Phys. Rev. B13, 299(1976) was useful to J. B.
 Goodenough, F. R. S. His paper, J. Phys. C 20, L789 (1987) on the
 microwave absorption, was useful to Nobel Laureate K. Alex M\"uller. 
His paper published in the Proc. Roy. Soc. A419, 287- 303(1988),
communicated by B. Bleaney, F.R.S. represents original work of interest to Nobel Laureate J. H. van Vleck.  He discovered the flux quantized energy levels in superconductors and the correct theory of 1/3 charge in quantum Hall effect}.
The correct theory of quantum Hall effect is given in ref.9.
\vskip1.25cm

\noindent{\bf5.~~References}
\begin{enumerate}
\item J. K. Jain, Phys. Rev. Lett. {\bf 63}, 199 (1989).
\item K. Park and J. K. Jain, {\bf 80}, 437 (1998)
\item S. S. Mandal, M. R. Peterson and J. K. Jain, Phys. Rev. Lett. {\bf 90}, 106403(2003).
\item G.'tHooft, Nuc. Phys. B {\bf 72},461 (1974);{\bf 75}, 461(1974); {\bf 79}. 276 (1974).
\item M. I. Dyakonov, cond-mat/0209206.
\item B. Farid, cond-mat/0003064.
\item W. Pan et al, Phys. Rev. Lett. {\bf 90}, 016801(2003)
\item K. N. Shrivastava, Bull. Am. Phys. Soc. J1.209 (2003)
\item K. N. Shrivastava, Introduction to quantum Hall effect,\\ 
      Nova Science Pub. Inc., N. Y. (2002).
\end{enumerate}
\vskip0.1cm
Note: Ref. 9 is available from:\\
 Nova Science Publishers, Inc.,\\
400 Oser Avenue, Suite 1600,\\
 Hauppauge, N. Y.. 11788-3619,\\
Tel.(631)-231-7269, Fax: (631)-231-8175,\\
 ISBN 1-59033-419-1 US$\$69$.\\
E-mail: novascience@Earthlink.net

\vskip5.5cm
\end{document}